\documentclass[sigconf,natbib=false,nonacm, screen]{acmart}
\usepackage[style=ieee,isbn=false,url=false,doi=false]{biblatex}
\usepackage{hyperref}
\usepackage{amsmath,amsfonts}
\usepackage{algorithmic}
\usepackage{graphicx}
\usepackage{textcomp}
\usepackage{booktabs}
\usepackage{multirow}
\usepackage{xcolor}
\usepackage{lipsum}
\usepackage{listings}
\usepackage{supertabular}
\usepackage{appendix}

\begin{document}

\title{ChatGPT and Human Synergy in Black-Box Testing: \\A Comparative Analysis}

\author{Hiroyuki Kirinuki}
\email{hiroyuki.kirinuki@ntt.com}
\affiliation{%
  \institution{NTT Software Innovation Center}
  \city{Tokyo}
  \country{Japan}
}
\author{Haruto Tanno}
\email{haruto.tanno@ntt.com}
\affiliation{%
  \institution{NTT Software Innovation Center}
  \city{Tokyo}
  \country{Japan}
}

\begin{abstract}
  In recent years, large language models (LLMs), such as ChatGPT, have been pivotal in advancing various artificial intelligence applications, including natural language processing and software engineering. A promising yet underexplored area is utilizing LLMs in software testing, particularly in black-box testing. This paper explores the test cases devised by ChatGPT in comparison to those created by human participants.
  In this study, ChatGPT (GPT-4) and four participants each created black-box test cases for three applications based on specifications written by the authors. The goal was to evaluate the real-world applicability of the proposed test cases, identify potential shortcomings, and comprehend how ChatGPT could enhance human testing strategies.
  ChatGPT can generate test cases that generally match or slightly surpass those created by human participants in terms of test viewpoint coverage. Additionally, our experiments demonstrated that when ChatGPT cooperates with humans, it can cover considerably more test viewpoints than each can achieve alone, suggesting that collaboration between humans and ChatGPT may be more effective than human pairs working together.
  Nevertheless, we noticed that the test cases generated by ChatGPT have certain issues that require addressing before use.
\end{abstract}

\maketitle

\section{Introduction}
The rise of large language models (LLMs) like ChatGPT marks a significant moment in artificial intelligence. LLM is not only applicable for tasks such as natural language processing, sentiment analysis, and automated customer support, but it also exhibits impressive versatility in the software engineering domain. For instance, developers are increasingly using LLM to instantly generate code in programming languages like Python and Java based on text descriptions, thereby boosting their productivity.

One of the key applications of LLM is in software testing, specifically in generating unit tests. Typically, this is referred to as white-box testing, where the tester is familiar with the software's internal structure, design, and implementation. In contrast, in black-box testing, the software's internal structure remains unknown, and testing is conducted based on requirements.

There has been some research on using LLMs for white-box test generation. Schäfer et al.~\cite{schafer_adaptive_2023} and Li et al.~\cite{li_finding_2023} have explored ChatGPT in this regard, demonstrating its potential for generating effective white-box test cases. However, using LLMs for black-box testing has not been delved into much.

Some studies have tried to tackle this, but they often focus on limited areas. For instance, Khaliq et al.~\cite{khaliq_transformers_2022} proposed using a transformer approach for GUI testing, where GPT-2 identified screens and produced Appium test cases. Jalil et al.~\cite{jalil_chatgpt_2023} looked at ChatGPT's ability to solve textbook problems for software testing education. Lastly, Liu et al. ~\cite{liu_chatting_2023} explored GPT-3's potential for human-like testing of mobile application GUIs.

While these studies are indeed intriguing, they do not clarify the capabilities of LLMs in black-box testing. Comparing black-box testing using LLM to that done by developers is crucial to genuinely assess its capabilities. Our paper focuses on shedding light on ChatGPT's abilities for black-box testing.

\section{Background and Related Work}

\subsection{Large language models}

LLMs such as GPT-3 and ChatGPT have significantly impacted various natural language processing tasks like text summarization, dialogue generation, and machine translation~\cite{brown_language_2020,zhang_perturbation_2020,zhang_extractive_2023,aharoni_massively_2019}. The foundational work on neural networks and language modeling leveraged substantial amounts of data and computational resources~\cite{bengio_neural_2000}.

LLMs predict the next words based on the preceding context, a factor that significantly contributes to their superior generalization ability. They surpass previous methods by a wide margin in text summarization and evaluation tasks, indicating a stronger alignment with human evaluations~\cite{liu_is_2023,fu_gptscore_2023,luo_chatgpt_2023}.
The success of many LLMs, such as BERT, GPT-2, and XLNet, is supported by the transformative self-attention mechanism of the Transformer model, which enables them to capture richer semantic and contextual relationships~\cite{vaswani_attention_2017, devlin_bert_2019,radford_language_nodate,yang_xlnet_2019}.

ChatGPT, developed by OpenAI, has notably influenced dialogue system development, excelling in both task-oriented and open-domain conversations~\cite{openai_gpt-4_2023,brown_language_2020}. Its ability to understand complex language patterns, generate coherent and diverse text, and generalize from previous contexts can also be leveraged for software testing.

\subsection{Black-Box Testing}

Black-box testing verifies the functionality of software without delving into its internal workings. It focuses on the inputs and outputs of the software system, working to detect errors and issues in the software's functionality. Test cases are designed based on the software's specifications and requirements, which makes it especially suitable for higher-level tests where the internal structure of the system under test is unknown.

Black-box testing employs a variety of techniques such as equivalence partitioning, boundary value analysis, decision table testing, and state transition testing to efficiently address system complexities. These methodologies are applicable in various phases of software testing, including unit, integration, system, acceptance, regression, and functional testing.

\begin{table*}[tb]
  \centering
  \caption{Descriptions of Experimental Applications}
  \label{tab:apps_and_descriptions}
  \begin{tabular}{l p{0.68\textwidth} l}
    \toprule
    Application               & Description                                                                                                                        & State     \\
    \midrule
    Password strength checker & Classifies the strength of a given password based on certain criteria, categorizing it into one of five levels.                    & Stateless \\
    Unit converter            & Converts values among different units across three categories: length, weight, and temperature.                                    & Stateless \\
    Budget planner            & Operable through seven commands, this tool aids users in logging various incomes and expenses to calculate the net profit or loss. & Stateful  \\
    \bottomrule
  \end{tabular}
\end{table*}

\subsection{Related Work}

Automating test case generation from system specifications is a central topic in software testing research. Generally, approaches in this area fall under three categories: those leveraging unified modeling language (UML) behavioral models, those utilizing natural language (NL) requirements, and those grounded in formal methods.

\subsubsection*{UML-based methodologies:}
Research in this area has largely focused on using UML behavioral models such as activity diagrams, statecharts, and sequence diagrams for test case generation~\cite{linzhang_generating_2004, briand_uml_2002, nayak_synthesis_2011,hasling_model_2008,samuel_slicing-based_2009,nebut_automatic_2006}. Notably, Nebut et al.~\cite{nebut_automatic_2006}, Gutierrez et al.~\cite{gutierrez_model-driven_2015}, and Briand and Labiche~\cite{briand_uml_2002} utilized system sequence diagrams and activity diagrams. Other significant works involved generating test cases from statecharts and NL scenarios~\cite{frohlich_automated_2000,santiago_junior_generating_2012}.

\subsubsection*{NL-based methodologies:}

This strand leverages natural language analysis techniques for automated test case generation, as seen in the works of Masuda et al.\cite{masuda_automatic_2016} and Leitao et al.~\cite{leitao_nlforspec_2007}. The latter introduced NLForSpec, a tool that translates NL test case descriptions into formal language, demonstrating a 91\% efficiency rate in mobile app testing.

\subsubsection*{Formal methods:}
These methods employ mathematical logic and set theory for creating clear system descriptions and behaviors, aiding in systematic test case derivation to ensure thorough coverage and correctness. Liu and Nakajima~\cite{liu_automatic_2022} introduced the ``V-Method'' for automated test case and oracle generation from formal specifications. Yang et al.~\cite{yang_formal_2013} focused on CTL* temporal logic specifications for deriving test cases, while Chang et al.~\cite{chang_test_2000} combined formal specifications with usage profiles to uncover subtle errors often overlooked by traditional techniques.

Our research explores the potential of large language models, especially ChatGPT, in creating test cases from specifications. We assess ChatGPT's efficiency against human-created test cases, aiming to identify its strengths and weaknesses to integrate it effectively into existing workflows.

\section{Evaluation}

We organized an experiment to gauge the capabilities of ChatGPT (GPT-4 in June 2023) in creating black-box test cases. Initially, both ChatGPT and four human perticipants were tasked with created test cases based on application specifications devised by the authors. The four participants have experience in program implementation, unit testing, and ad-hoc testing, but were not experts in test design for black-box testing. We then extracted the test viewpoints contained in the test cases and performed a comparative analysis.

In the context of this study, a ``test viewpoint'' is conceptualized as a specific aspect or criterion that a single test case aims to validate. This can be understood as the unique perspective from which a particular test case examines the system under test. For example, a test case might be designed to assess system behavior when encountering multibyte or unicode characters, or to determine the system's response to input that exceeds maximum string length parameters. Each test case ideally encapsulates a single test viewpoint to ensure clarity in identifying potential bugs or system faults. Incorporating multiple viewpoints within a single test case can obscure the root causes of any issues uncovered, thereby complicating the debugging process. Given the nature of black-box testing, we determined that evaluating test cases based on their test viewpoints is more appropriate and effective, rather than relying on code coverage metrics.

The primary focuses of this analysis were to:

\begin{itemize}
  \item Assess the real-world applicability of test cases suggested by ChatGPT.
  \item Identify test viewpoints that ChatGPT often overlooks when creating test cases compared to human testers, and vice versa.
  \item Find out how human testers might leverage ChatGPT to augment their testing approaches.
\end{itemize}

For this experiment, we chose three applications as subjects: Password strength checker, Unit converter, and Budget planner. Among these, the Password strength checker and Unit converter are stateless, while the Budget planner is statefull.
These applications were specified by the authors as command-line based applications, but not actually implemented. This is because these applications do not need to be available in our experiments. A detailed description of each application can be found in Table~\ref{tab:apps_and_descriptions}. We have shared the application specifications, ChatGPT prompts, the resulting test cases, and test viewpoints extracted from the test cases on \url{https://zenodo.org/records/10476924}.

\subsection{Test Case Generation by ChatGPT}
\begin{figure}
  \centering
  \includegraphics[width=0.90\columnwidth]{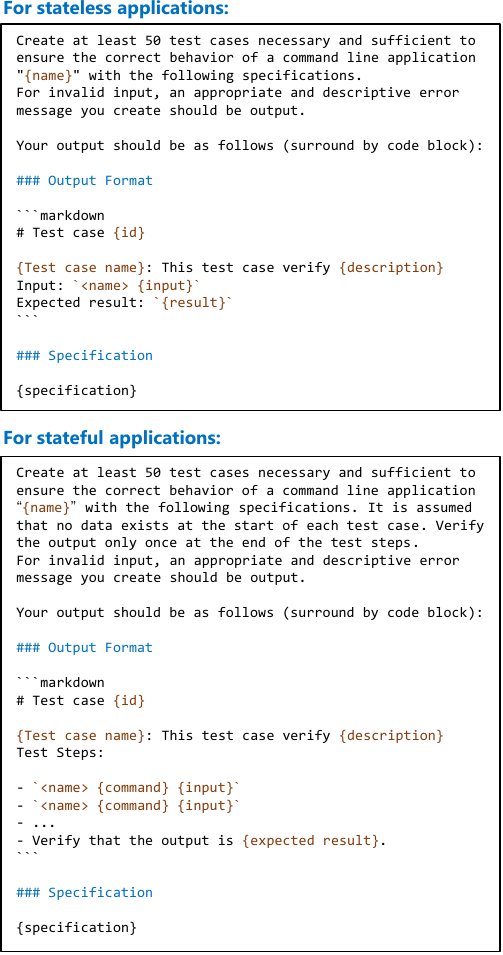}
  \caption{Prompt given to ChatGPT}
  \label{tab:prompt}
  \Description{Prompt given to ChatGPT}
\end{figure}

We provided ChatGPT with the application specifications and instructed it to generate test cases based on these specifications.
Figure \ref{tab:prompt} shows the prompt given to ChatGPT for the creation of the test suite.
For stateless applications, we asked ChatGPT to suggest input values and their expected results. For the stateful application, we guided ChatGPT to generate the testing procedures and then indicate the anticipated outcomes.
ChatGPT was instructed to output the name of the test case and what it validates. We do not provide ChatGPT with any knowledge for test design or testing techniques to be applied. We also instructed ChatGPT to create a total of 50 test cases for each application specification in one session. We chose this number of test cases because we judged it sufficient to comprehensively test the target application, as indicated by our preliminary experiments. These experiments also showed that ChatGPT could not determine on its own when enough test cases had been generated.

\begin{figure}
  \centering
  \includegraphics[width=0.90\columnwidth]{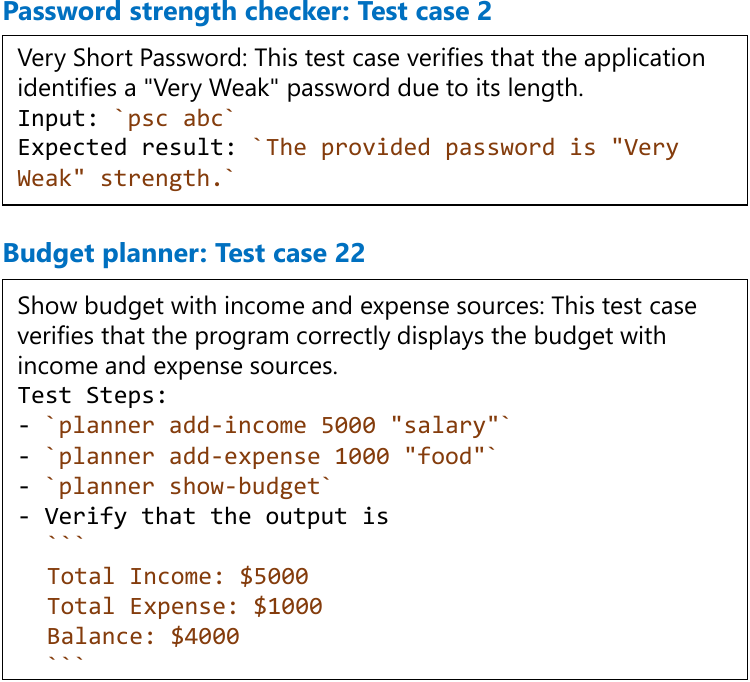}
  \caption{Examples of test cases generated by ChatGPT}
  \label{fig:example_testcase}
  \Description{Examples of test cases generated by ChatGPT}
\end{figure}

The examples of the produced test cases are showed in Figure \ref{fig:example_testcase}, with the stateless application ``Password strength checker'' delineating the input values and expected results, and the stateful ``Budget planner'' depicting the testing procedures along with the expected results.

\subsection{Test Viewpoint Analysis}
There were five test suites in total, each being a collection of test cases for a single application, derived from the test cases generated by four participants and ChatGPT. The authors established ``basic viewpoints'' prior to this experiment to validate the essential functionalities of the three applications. We examined these test suites closely and extracted test viewpoints shown in any of the test cases. We then evaluated how many of these viewpoints were covered by each test suite. The basic and extracted viewpoints and evaluation results were reviewed and revised by the authors and the four participants.

Out of all the test viewpoints, those included by at least two of the five test suites were considered worthy for testing and were thus labeled as ``effective viewpoints''. Consequently, every basic viewpoint we established was classified as an effective viewpoint. Regarding the overall count of test viewpoints, the breakdown is as follows: Password strength checker with 36, Unit converter with 29, and Budget planner with 41. From these, the count of effective viewpoints was 24, 23, and 31, respectively.
Details on which test viewpoints were extracted for each test suite are provided in the Appendix.

\begin{table*}[tb]
  \centering
  \caption{The number of covered effective viewpoints by each test suite}
  \label{tab:covered_viewpoints}
  \begin{tabular}{llcccccccccc}
    \toprule
    \multirow{2.5}{*}{Application}             & \multirow{2.5}{*}{Viewpoint type} & \multicolumn{9}{c}{\texttt{\#} of covered viewpoints} & \multirow{2.5}{*}{\shortstack[c]{\texttt{\#} of effective                                         \\viewpoints}} \\
    \cmidrule(lr){3-11}
                                               &                                   & ChatGPT                                               & A                                                         & B  & C  & D  & A+ & B+ & C+ & D+ &    \\
    \midrule
    \multirow{3}{*}{Password strength checker} & Basic                             & 10                                                    & 10                                                        & 12 & 10 & 10 & 12 & 12 & 11 & 13 & 13 \\
                                               & Extracted                         & 9                                                     & 5                                                         & 8  & 9  & 10 & 10 & 10 & 11 & 11 & 11 \\
                                               & All                               & 19                                                    & 15                                                        & 20 & 19 & 20 & 22 & 22 & 22 & 24 & 24 \\
    \midrule
    \multirow{3}{*}{Unit converter}            & Basic                             & 2                                                     & 2                                                         & 2  & 2  & 2  & 2  & 2  & 2  & 2  & 2  \\
                                               & Extracted                         & 17                                                    & 13                                                        & 16 & 16 & 16 & 20 & 18 & 20 & 21 & 21 \\
                                               & All                               & 19                                                    & 15                                                        & 18 & 18 & 18 & 22 & 20 & 22 & 23 & 23 \\
    \midrule
    \multirow{3}{*}{Budget planner}            & Basic                             & 7                                                     & 7                                                         & 7  & 7  & 7  & 7  & 7  & 7  & 7  & 7  \\
                                               & Extracted                         & 17                                                    & 18                                                        & 16 & 16 & 21 & 23 & 19 & 22 & 22 & 24 \\
                                               & All                               & 24                                                    & 25                                                        & 23 & 23 & 28 & 30 & 26 & 29 & 29 & 31 \\
    \midrule
    \multirow{1}{*}{}                          & Total                             & 62                                                    & 55                                                        & 61 & 60 & 66 & 74 & 68 & 73 & 76 & 78 \\
    \bottomrule
  \end{tabular}
\end{table*}

\begin{figure*}[tb]
  \centering
  \includegraphics[width=0.85\textwidth]{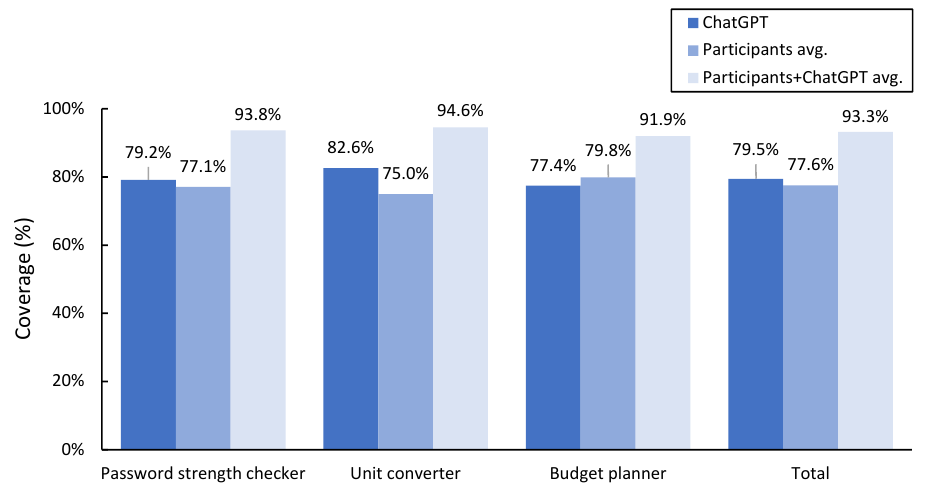}
  \caption{Coverage of effective viewpoints}
  \label{fig:coverage}
  \Description{Coverage of test viewpoints}
\end{figure*}

\subsection{Result}
Table \ref{tab:covered_viewpoints} displays the number of effective viewpoints covered by the test suites created by ChatGPT and the four participants, labeled as A through D. During this experiment, we also assessed the possible coverage of effective viewpoints if participants had referred to ChatGPT's test suite. Entries A+ to D+ in the table illustrate the number of covered effective viewpoints when assuming that participants A to D consulted ChatGPT. This number is calculated from the union of the viewpoints covered both by ChatGPT and each individual participant. The basic assumption is that participants would recognize and adopt the overlooked viewpoints presented by ChatGPT.

Figure \ref{fig:coverage} compares the average test viewpoint coverage between ChatGPT and the participants. In two of the three applications, ChatGPT achieved slightly better coverage, while for the other one, the participants exhibited superior performance.
By collaborating with ChatGPT, the participants achieved a markedly higher average coverage, reaching 93.3\% of the effective viewpoints. Additionally, participants took an average of 198 minutes to construct the test suites for the three applications.

\section{Discussion}
Although the participants were not experts in test design, the test suites created by ChatGPT were generally on par with or slightly superior to those created by the participants. Considering the time required for manual test design, we believe ChatGPT's performance is practical. A key takeaway from this experiment is that utilizing ChatGPT can help in reducing the number of overlooked test viewpoints during the creation of black-box tests.

\subsection{Limitations of ChatGPT}
The experiment also highlighted some points of caution when using ChatGPT. The first issue is that ChatGPT often misses test viewpoints associated with boundary and maximum values; about half of the overlooked viewpoints in this experiment pertained to these aspects. Therefore, when developers use ChatGPT in their test design, they should either provide additional instructions to ChatGPT or carefully cover these test viewpoints.

The second issue to note is the occasional mismatch between the test case descriptions formulated by ChatGPT and the respective input values, test procedures, and expected outcomes. In instances where such inconsistencies emerged during this study, we presumed the test case description to be correct. This difficulty is somewhat tied to the first issue, but it frequently occurred that ChatGPT misinterpreted the required length of strings in the tests. For example, ChatGPT suggested strings exceeding ten characters when the requirement was for eight or nine characters. Therefore, it is advisable not to depend solely on the input-output values or test procedures produced by ChatGPT when performing black-box tests.

The third point of consideration is ChatGPT's inability to produce a large batch of test cases at once. During the study, we directed ChatGPT to create 50 test cases for a single application directive; however, as it approached 40 test case, it exhibited a tendency to forget the application specification or to suggest tests identical to previously proposed ones. This restriction might stem from the limited number of tokens it can process at a time. While future LLM advancements might alleviate this, for more intricate test targets than those used in this study, it is necessary to devise a way to divide the specifications well and generate test cases for each of them.

\subsection{Comparison of ChatGPT and Human Characteristics}

As mentioned previously, while ChatGPT shows weaknesses in handling boundary and maximum/minimum values, no significant differences were observed in the test viewpoints covered by ChatGPT compared to humans. Nevertheless, ChatGPT is comparable or superior coverage, despite missing certain test viewpoints, suggests proficiency in other areas.

The experimental results indicated that humans working alongside ChatGPT covered more test viewpoints than humans working independently. However, this does not imply that a human--ChatGPT pair is superior to a human--human pair. To further analyze this, we evaluated the similarity of test viewpoints confirmed by human--ChatGPT pairs against those confirmed by human--human pairs. The test viewpoints from each test suite were treated as sets, and their similarities were calculated using the Jaccard index and Cosine similarity.

The Jaccard index, defined as the ratio of the intersection to the union of two sets, is mathematically represented as:
\[ J(X, Y) = \frac{|X \cap Y|}{|X \cup Y|} \]
In this context, for human--ChatGPT pairs, it was calculated as:
\[ J_{\text{human--ChatGPT}} = \frac{\text{Avg}(|\text{A} \cap \text{ChatGPT}| + \cdots + |\text{D} \cap \text{ChatGPT}|)}{\text{Avg}(|\text{A} \cup \text{ChatGPT}| + \cdots + |\text{D} \cup \text{ChatGPT}|)} \]
Similarly, for human--human pairs, the calculation included all six possible pair combinations from A to D.

Cosine similarity was calculated by treating each test viewpoint as a vector element, assigning 1 if covered by a test suite and 0 otherwise. The formula for Cosine similarity is:
\[ \text{Cosine}(X, Y) = \frac{X \cdot Y}{||X|| \times ||Y||} \]
where \( X \cdot Y \) represents the dot product of vectors X and Y, and \( ||X|| \) and \( ||Y|| \) are the magnitudes of vectors X and Y, respectively.
Both indices, the Jaccard index and Cosine similarity, range from 0 to 1, with values closer to 1 indicating higher similarity.

As shown in Table \ref{tab:viewpoint_similarity}, ChatGPT (GPT) and human (HMN) pairs exhibit lower similarity in test viewpoints compared to human-human pairs. This suggests that collaborations between ChatGPT and humans might cover a broader range of test viewpoints than human pairs alone.

\begin{table}[tb]
  \centering
  \caption{Similarity of covered test viewpoints}
  \label{tab:viewpoint_similarity}
  \begin{tabular}{@{}cccc@{}}
    \toprule
    \multicolumn{2}{c}{Jaccard index} & \multicolumn{2}{c}{Cosine Similarity}                             \\
    \cmidrule(r){1-2} \cmidrule(l){3-4}
    HMN--GPT                          & HMN--HMN                              & HMN--GPT       & HMN--HMN \\
    \textbf{0.565}                    & 0.577                                 & \textbf{0.724} & 0.733    \\
    \bottomrule
  \end{tabular}
\end{table}

\section{Conclusion}
In this study, we explored the black-box test design capabilities of the current ChatGPT (GPT-4). The results suggested that ChatGPT can generate test cases equivalent or superior to those created by humans, hinting at the possibility of enhanced test viewpoint coverage through human collaboration. Furthermore, it indicated that collaboration between ChatGPT and humans could cover a broader range of test viewpoints compared to human-only collaboration.

However, challenges such as ChatGPT overlooking test viewpoints related to boundary values or maximum/minimum values were also identified. Based on these findings, we plan to tackle these challenges in our future work. Our primary task is to ensure ChatGPT does not overlook any commonly missed test viewpoints. Determining whether prompt engineering can mitigate this issue is a crucial step.

Subsequently, we need to develop a feasible test process utilizing ChatGPT. Given the potential discrepancies between ChatGPT's test case descriptions, associated inputs, processes, and expected outcomes, as well as batch size limitations, considerable practical challenges arise. To apply this practically, we need to find ways to mitigate these issues. Additionally, though not explored in this study, assessing the uniformity of ChatGPT's output is vital, investigating whether it proposes diverse or similar test cases in each run to establish effective usage strategies.

By overcoming these challenges, we aim to enable non-testing experts to prform black-box testing quickly and efficiently, surpassing testing experts. Additionally, we envision using ChatGPT to create tests for more complex applications, like web applications, and for flexible testing approaches, such as exploratory testing.

\printbibliography

\appendix

\begin{appendix}

  \section*{Appendix: Test Viewpoints in Each Test Suite}
  \begin{table*}[htbp]
    \centering
    \caption{Extracted test viewpoints from Password Strength Checker}
    \begin{tabular}{@{}p{3cm} p{8.7cm} cccccc@{}}
      \toprule
      Category & Test Viewpoint                                                                                                                                                          & ChatGPT    & A          & B          & C          & D          & \shortstack{Effective \\ Viewpoint} \\
      \midrule
      \multirow{3}{*}{\shortstack[l]{Basic Viewpoint:                                                                                                                                                                                                                             \\ Very Weak judgment}}
               & 7 characters or fewer \& 1 type of character                                                                                                                            & \checkmark & \checkmark & \checkmark & \checkmark &            & \checkmark            \\
               & 7 characters or fewer \& 2 or more types of characters                                                                                                                  &            & \checkmark & \checkmark &            & \checkmark & \checkmark            \\
               & 8 characters or more \& 1 type of character                                                                                                                             & \checkmark & \checkmark & \checkmark & \checkmark & \checkmark & \checkmark            \\
      \midrule
      \multirow{3}{*}{\shortstack[l]{Basic Viewpoint:                                                                                                                                                                                                                             \\ Weak judgment}}
               & 8,9 characters \& 2 or 3 types of characters (excluding symbols)                                                                                                        & \checkmark & \checkmark & \checkmark & \checkmark & \checkmark & \checkmark            \\
               & 8,9 characters \& 2 or more types of characters including symbols                                                                                                       & \checkmark &            & \checkmark & \checkmark & \checkmark & \checkmark            \\
               & 10 characters or more \& 2 or 3 types of characters (excluding symbols)                                                                                                 & \checkmark & \checkmark & \checkmark & \checkmark & \checkmark & \checkmark            \\
      \midrule
      \multirow{3}{*}{\shortstack[l]{Basic Viewpoint:                                                                                                                                                                                                                             \\ Medium judgment}}
               & 10,11 characters \& 2 types of characters including symbols                                                                                                             & \checkmark & \checkmark & \checkmark & \checkmark &            & \checkmark            \\
               & 10,11 characters \& 3 or more types of characters including symbols                                                                                                     & \checkmark &            & \checkmark &            & \checkmark & \checkmark            \\
               & 12 characters or more \& 2 types of characters including symbols                                                                                                        &            & \checkmark &            & \checkmark & \checkmark & \checkmark            \\
      \midrule
      \multirow{3}{*}{\shortstack[l]{Basic Viewpoint:                                                                                                                                                                                                                             \\ Strong judgment}}
               & 12–15 characters \& 3 types of characters including symbols                                                                                                             & \checkmark & \checkmark & \checkmark & \checkmark &            & \checkmark            \\
               & 12–15 characters \& 4 types of characters                                                                                                                               &            &            & \checkmark &            & \checkmark & \checkmark            \\
               & 16 characters or more \& 3 types of characters including symbols                                                                                                        & \checkmark & \checkmark & \checkmark & \checkmark & \checkmark & \checkmark            \\
      \midrule
      Basic Viewpoint:                                                                                                                                             Very Strong judgment
               & 16 characters or more \&                                                                                                                          4 types of characters & \checkmark & \checkmark & \checkmark & \checkmark & \checkmark & \checkmark            \\
      \midrule
      \multirow{2}{*}{Boundary Value}
               & String length boundary values (excluding minimum and maximum lengths)                                                                                                   &            &            & \checkmark & \checkmark & \checkmark & \checkmark            \\
      \midrule
      \multirow{2}{*}{Minimum/Maximum}
               & Minimum length (1)                                                                                                                                                      &            &            &            & \checkmark &            &                       \\
               & Maximum length (100)                                                                                                                                                    &            & \checkmark &            & \checkmark & \checkmark & \checkmark            \\
      \midrule
      \multirow{4}{*}{\shortstack[l]{Character Type                                                                                                                                                                                                                               \\ Combinations}}
               & 4 ways to choose 1 type of character                                                                                                                                    & \checkmark &            & \checkmark & \checkmark & \checkmark & \checkmark            \\
               & 3 combinations of 2 types of characters excluding symbols                                                                                                               & \checkmark &            & \checkmark & \checkmark & \checkmark & \checkmark            \\
               & 3 combinations of 2 types of characters including symbols                                                                                                               & \checkmark &            & \checkmark & \checkmark & \checkmark & \checkmark            \\
               & 3 combinations of 3 types of characters including symbols                                                                                                               & \checkmark &            & \checkmark & \checkmark & \checkmark & \checkmark            \\
      \midrule
      \multirow{5}{*}{Inappropriate Strings}
               & Failing by not providing a string                                                                                                                                       & \checkmark & \checkmark & \checkmark &            & \checkmark & \checkmark            \\
               & Failing by providing multiple strings                                                                                                                                   & \checkmark & \checkmark &            &            & \checkmark & \checkmark            \\
               & Failing by providing a string longer than the maximum length                                                                                                            & \checkmark & \checkmark & \checkmark & \checkmark & \checkmark & \checkmark            \\
               & Failing by providing disallowed symbols                                                                                                                                 & \checkmark &            &            &            &            &                       \\
               & Failing by providing multibyte/unicode characters                                                                                                                       & \checkmark & \checkmark & \checkmark & \checkmark & \checkmark & \checkmark            \\
      \midrule
      \multirow{10}{*}{Others}
               & Linux command line “pipe”                                                                                                                                               &            & \checkmark &            &            &            &                       \\
               & Repeating the same character                                                                                                                                            & \checkmark &            &            &            &            &                       \\
               & Randomly scattering types of characters evenly                                                                                                                          & \checkmark &            &            &            &            &                       \\
               & All kinds of lowercase letters                                                                                                                                          &            &            &            & \checkmark &            &                       \\
               & All kinds of uppercase letters                                                                                                                                          &            &            &            & \checkmark &            &                       \\
               & All digits                                                                                                                                                              &            &            &            & \checkmark &            &                       \\
               & Using all symbols                                                                                                                                                       & \checkmark &            &            & \checkmark &            & \checkmark            \\
               & Testing the alphabet in reverse order                                                                                                                                   & \checkmark &            &            &            &            &                       \\
               & Testing invisible unicode characters                                                                                                                                    & \checkmark &            &            &            &            &                       \\
               & Failing by providing spaces at the beginning and end                                                                                                                    &            & \checkmark &            &            &            &                       \\
      \bottomrule
    \end{tabular}
  \end{table*}

  \begin{table*}[htbp]
    \centering
    \caption{Extracted test viewpoints from Unit Converter}
    \begin{tabular}{@{}lp{8.5cm}cccccc@{}}
      \toprule
      Category                                & Test Viewpoint                                                                                                & ChatGPT    & A          & B          & C          & D          & \shortstack{Effective \\ Viewpoint} \\
      \midrule
      \multirow{4}{*}{Basic Viewpoint}        & For all units of length, appearing in either source or target and successfully converting                     & \checkmark & \checkmark & \checkmark & \checkmark & \checkmark & \checkmark            \\
                                              & For all units of temperature, appearing in either source or target and successfully converting                & \checkmark & \checkmark & \checkmark & \checkmark & \checkmark & \checkmark            \\
      \midrule
      \multirow{4}{*}{Combination}            & Exhaustive coverage of two-unit combinations for length                                                       & \checkmark &            & \checkmark & \checkmark & \checkmark & \checkmark            \\
                                              & All units of length appear in both source and target                                                          & \checkmark & \checkmark & \checkmark &            & \checkmark & \checkmark            \\
                                              & Exhaustive coverage of two-unit combinations for temperature                                                  & \checkmark & \checkmark & \checkmark & \checkmark & \checkmark & \checkmark            \\
                                              & All units of temperature appear in both source and target                                                     & \checkmark & \checkmark & \checkmark &            & \checkmark & \checkmark            \\
      \midrule
      \multirow{3}{*}{Unit Error}             & Conversion to a unit from a different category                                                                & \checkmark & \checkmark & \checkmark & \checkmark & \checkmark & \checkmark            \\
                                              & Conversion between the same units                                                                             & \checkmark &            & \checkmark & \checkmark & \checkmark & \checkmark            \\
                                              & Providing a length unit that is not supported                                                                 & \checkmark &            & \checkmark &            & \checkmark & \checkmark            \\
      \midrule
      \multirow{2}{*}{\shortstack[l]{Value Variation                                                                                                                                                                                                   \\ (Common)}} & When the pre-conversion value is an integer                                                                   & \checkmark & \checkmark & \checkmark & \checkmark & \checkmark & \checkmark            \\
                                              & When the pre-conversion value is a decimal                                                                    & \checkmark & \checkmark & \checkmark & \checkmark & \checkmark & \checkmark            \\
      \midrule
      \multirow{3}{*}{\shortstack{Value Error                                                                                                                                                                                                          \\ (Common)}} & Providing a value exceeding the maximum value                                                                 & \checkmark & \checkmark & \checkmark & \checkmark & \checkmark & \checkmark            \\
                                              & Providing a value up to the third decimal place                                                               &            & \checkmark & \checkmark & \checkmark & \checkmark & \checkmark            \\
                                              & Providing an invalid character in the pre-conversion value                                                    & \checkmark & \checkmark &            & \checkmark & \checkmark & \checkmark            \\
      \midrule
      \multirow{2}{*}{Length Error}           & When the pre-conversion value for length is negative                                                          & \checkmark & \checkmark & \checkmark & \checkmark &            & \checkmark            \\
                                              & When the pre-conversion value for length is zero                                                              &            & \checkmark &            & \checkmark & \checkmark & \checkmark            \\
      \midrule
      \multirow{4}{*}{Temperature Variation } & Successful conversion when the pre-conversion value in Celsius/Fahrenheit is zero                             &            &            &            & \checkmark &            &                       \\
                                              & Successful conversion when the pre-conversion value in Celsius/Fahrenheit is negative                         & \checkmark &            &            &            &            &                       \\
      \midrule
      \multirow{4}{*}{Temperature Error}      & When the pre-conversion value in Kelvin is zero                                                               & \checkmark &            &            &            & \checkmark & \checkmark            \\
                                              & When the pre-conversion value in Kelvin is negative                                                           & \checkmark & \checkmark & \checkmark & \checkmark &            & \checkmark            \\
                                              & Giving a value smaller than the minimum for Celsius/Fahrenheit (resulting in negative Kelvin post-conversion) &            & \checkmark &            &            & \checkmark & \checkmark            \\
      \midrule
      \multirow{4}{*}{Argument Format Error}  & When no value is provided                                                                                     & \checkmark &            & \checkmark & \checkmark &            & \checkmark            \\
                                              & When no 'from' is provided                                                                                    & \checkmark &            & \checkmark & \checkmark &            & \checkmark            \\
                                              & When no 'to' is provided                                                                                      & \checkmark &            & \checkmark & \checkmark &            & \checkmark            \\
                                              & When multiple inputs are made for from, value, and to                                                         &            &            &            &            & \checkmark &                       \\
      \midrule
      \multirow{6}{*}{Maximum/Minimum}        & Providing the minimum value for length/Kelvin (0.01)                                                          &            &            &            &            & \checkmark &                       \\
                                              & Providing the maximum value (1000)                                                                            &            &            &            & \checkmark & \checkmark & \checkmark            \\
                                              & Giving the minimum Celsius value (resulting in 0 Kelvin post-conversion)                                      & \checkmark &            &            &            &            &                       \\
                                              & Giving the minimum Fahrenheit value (resulting in 0 Kelvin post-conversion)                                   & \checkmark &            &            &            &            &                       \\
      \bottomrule
    \end{tabular}
  \end{table*}

  \begin{table*}[htbp]
    \centering
    \caption{Extracted test viewpoints from Budget Planner}
    \begin{tabular}{@{}lp{7.5cm}cccccc@{}}
      \toprule
      Category                                                                                                                                                                                                          & Test Viewpoint                                                                        & ChatGPT    & A          & B          & C          & D          & \shortstack{Effective \\ Viewpoint} \\
      \midrule
      \multirow{7}{*}{Basic Viewpoint}                                                                                                                                                                                  & Successfully add an income                                                            & \checkmark & \checkmark & \checkmark & \checkmark & \checkmark & \checkmark            \\
                                                                                                                                                                                                                        & Successfully add an expense                                                           & \checkmark & \checkmark & \checkmark & \checkmark & \checkmark & \checkmark            \\
                                                                                                                                                                                                                        & Successfully remove an income                                                         & \checkmark & \checkmark & \checkmark & \checkmark & \checkmark & \checkmark            \\
                                                                                                                                                                                                                        & Successfully remove an expense                                                        & \checkmark & \checkmark & \checkmark & \checkmark & \checkmark & \checkmark            \\
                                                                                                                                                                                                                        & Successfully display incomes                                                          & \checkmark & \checkmark & \checkmark & \checkmark & \checkmark & \checkmark            \\
                                                                                                                                                                                                                        & Successfully display expenses                                                         & \checkmark & \checkmark & \checkmark & \checkmark & \checkmark & \checkmark            \\
                                                                                                                                                                                                                        & Successfully display the budget                                                       &            & \checkmark &            & \checkmark & \checkmark & \checkmark            \\
      \midrule
      \multirow{8}{*}{Amount Input}                                                                                                                                                                                     & Giving 0                                                                              & \checkmark & \checkmark & \checkmark & \checkmark & \checkmark & \checkmark            \\
                                                                                                                                                                                                                        & Giving a negative number                                                              &            &            & \checkmark & \checkmark & \checkmark & \checkmark            \\
                                                                                                                                                                                                                        & Not providing an amount                                                               & \checkmark & \checkmark & \checkmark & \checkmark & \checkmark & \checkmark            \\
                                                                                                                                                                                                                        & Giving a decimal number                                                               & \checkmark & \checkmark & \checkmark & \checkmark & \checkmark & \checkmark            \\
                                                                                                                                                                                                                        & Giving a value exceeding the maximum value                                            &            & \checkmark & \checkmark & \checkmark & \checkmark & \checkmark            \\
                                                                                                                                                                                                                        & Giving a non-numeric value                                                            &            & \checkmark &            &            &            &                       \\
                                                                                                                                                                                                                        & Giving a multibyte/Unicode character                                                  &            &            &            &            & \checkmark &                       \\
                                                                                                                                                                                                                        & Giving multiple amounts                                                               & \checkmark & \checkmark & \checkmark & \checkmark & \checkmark & \checkmark            \\
      \midrule
      \multirow{8}{*}{Source Input}                                                                                                                                                                                     & Not providing a source                                                                & \checkmark &            &            &            & \checkmark & \checkmark            \\
                                                                                                                                                                                                                        & Giving an empty string                                                                & \checkmark & \checkmark & \checkmark & \checkmark & \checkmark & \checkmark            \\
                                                                                                                                                                                                                        & Exceeding the maximum string length                                                   &            &            &            & \checkmark &            &                       \\
                                                                                                                                                                                                                        & Giving a non-string value                                                             &            & \checkmark &            &            &            &                       \\
                                                                                                                                                                                                                        & Giving a multibyte/Unicode character                                                  & \checkmark &            &            &            &            &                       \\
                                                                                                                                                                                                                        & Including a space                                                                     &            & \checkmark &            &            & \checkmark & \checkmark            \\
                                                                                                                                                                                                                        & Not enclosing in double quotations                                                    &            &            &            &            & \checkmark &                       \\
                                                                                                                                                                                                                        & Giving multiple sources                                                               & \checkmark & \checkmark & \checkmark &            & \checkmark & \checkmark            \\
      \midrule
      \multirow{9}{*}{Internal State}                                                                                                                                                                                   & Exceeding the maximum number of registrations                                         & \checkmark & \checkmark & \checkmark & \checkmark & \checkmark & \checkmark            \\
                                                                                                                                                                                                                        & Deleting a non-existent source                                                        & \checkmark & \checkmark & \checkmark &            & \checkmark & \checkmark            \\
                                                                                                                                                                                                                        & Displaying when there is no source                                                    & \checkmark & \checkmark & \checkmark & \checkmark & \checkmark & \checkmark            \\
                                                                                                                                                                                                                        & Registering multiple sources                                                          & \checkmark & \checkmark & \checkmark & \checkmark & \checkmark & \checkmark            \\
                                                                                                                                                                                                                        & Verifying that deletion is reflected in the state                                     & \checkmark & \checkmark & \checkmark & \checkmark & \checkmark & \checkmark            \\
                                                                                                                                                                                                                        & Registering with the same name for income or expense                                  &            &            &            &            & \checkmark &                       \\
                                                                                                                                                                                                                        & Registering an expense source with the same name as an income source (and vice versa) & \checkmark &            &            &            &            &                       \\
                                                                                                                                                                                                                        & Deleting an income source as an expense (and vice versa)                              &            & \checkmark &            & \checkmark &            & \checkmark            \\
      \midrule
      \multirow{3}{*}{Maximum                                                                                                                                                                             Value}        & Maximum value for amount                                                              &            & \checkmark &            & \checkmark &            & \checkmark            \\
                                                                                                                                                                                                                        & Maximum string length                                                                 &            & \checkmark &            &            & \checkmark & \checkmark            \\
                                                                                                                                                                                                                        & Maximum number of registrations                                                       & \checkmark & \checkmark & \checkmark & \checkmark &            & \checkmark            \\
      \midrule
      \multirow{6}{*}{Budget                                                                                                                                                                               Calculation} & Having both income and expenses                                                       & \checkmark &            &            &            & \checkmark & \checkmark            \\
                                                                                                                                                                                                                        & Income only                                                                           & \checkmark &            &            &            & \checkmark & \checkmark            \\
                                                                                                                                                                                                                        & Expenses only                                                                         & \checkmark &            & \checkmark &            & \checkmark & \checkmark            \\
                                                                                                                                                                                                                        & Having neither income nor expenses                                                    & \checkmark &            & \checkmark & \checkmark & \checkmark & \checkmark            \\
                                                                                                                                                                                                                        & Verifying both positive and negative balances                                         &            &            &            &            & \checkmark &                       \\
                                                                                                                                                                                                                        & Having both income and expenses with the same amount                                  &            &            & \checkmark &            &            &                       \\
      \bottomrule
    \end{tabular}
  \end{table*}
\end{appendix}

\end{document}